\documentclass[aps,pre,twocolumn,amsmath,showpacs,amssymb,superscriptaddress]{revtex4}
\usepackage{graphicx}
\usepackage{color}

\begin{document}
\title{Ambiguities in recurrence-based complex network representations of time series}
\author{Reik V. Donner}
    \email{donner@vwi.tu-dresden.de}
    \affiliation{Max Planck Institute for Physics of Complex Systems, N\"othnitzer Str.~38, 01187 Dresden, Germany}
    \affiliation{Potsdam Institute for Climate Impact Research, P.O. Box 601203, 14412 Potsdam, Germany}
    \affiliation{Institute for Transport and Economics, Dresden University of Technology, W\"urzburger Str.~35, 01187 Dresden, Germany}
\author{Yong Zou}
    \email{yong.zou@pik-potsdam.de}
    \affiliation{Potsdam Institute for Climate Impact Research, P.O. Box 601203, 14412 Potsdam, Germany}
\author{Jonathan F. Donges}
    \affiliation{Potsdam Institute for Climate Impact Research, P.O. Box 601203, 14412 Potsdam, Germany}
    \affiliation{Department of Physics, Humboldt University Berlin, Newtonstr.~15, 12489 Berlin, Germany}
\author{Norbert Marwan}
    \affiliation{Potsdam Institute for Climate Impact Research, P.O. Box 601203, 14412 Potsdam, Germany}
\author{J\"urgen Kurths}
 \affiliation{Potsdam Institute for Climate Impact Research, P.O. Box 601203, 14412 Potsdam, Germany}
    \affiliation{Department of Physics, Humboldt University Berlin, Newtonstr.~15, 12489 Berlin, Germany}

\date{\today}

\begin{abstract}
Recently, different approaches have been proposed for studying basic properties of time series from a complex network perspective. In this work, the corresponding potentials and limitations of networks based on recurrences in phase space are investigated in some detail. We discuss the main requirements that permit a feasible system-theoretic interpretation of network topology in terms of dynamically invariant phase space properties. Possible artifacts induced by disregarding these requirements are pointed out and systematically studied. Finally, a rigorous interpretation of the clustering coefficient and the betweenness centrality in terms of invariant objects is proposed.
\end{abstract}

\maketitle

During the last decade, increasing interest has arisen in structural and dynamical properties of complex networks \cite{NetworkReviews}. Particular efforts have been spent on reconstructing network topologies from experimental data, e.g., in ecology \cite{Montoya2006}, social systems \cite{Freeman1979}, neuroscience \cite{Zhou2006}, or atmospheric dynamics \cite{Donges2009}. The latter example yields a complex network representation of a continuous system, which suggests that applying a similar spatial discretization to the phase space of dynamical systems and using complex network methods as a novel tool for time series analysis could be feasible, too \cite{Zhang2006,Marwan2009}. For this purpose, different methods have been proposed (for a comparative review, see \cite{Donner2009}) and successfully applied to real-world as well as model systems.

Many existing methods for transforming time series into complex network representations have in common that they define the connectivity of a complex network -- similar to the spatio-temporal case -- by the mutual proximity of different parts (e.g., individual states, state vectors, or cycles) of a single trajectory. In this work, we particularly consider \textit{recurrence networks}, which are based on the concept of recurrences in phase space \cite{Eckmann1987,Marwan2007} and provide a generic way for analyzing phase space properties in terms of network topology \cite{Donner2009,Xu2008,Gao2009,Marwan2009}. Here, the basic idea is to interpret the recurrence matrix
\begin{equation}
R_{i,j}(\varepsilon)=\Theta(\varepsilon-\|\mathbf{x}_i-\mathbf{x}_j \|)
\end{equation}
\noindent
associated with a dynamical system's trajectory, i.e., a binary matrix that encodes whether or not the phase space distance between two observed ``states'' $\mathbf{x}_i$ and $\mathbf{x}_j$ is smaller than a certain recurrence threshold $\varepsilon$, as the adjacency matrix of an undirected complex network. Since a single finite-time trajectory may however not necessarily represent the typical long-term behavior of the underlying system, the resulting network properties may depend -- among others -- on the length $N$ of the considered time series (i.e., the network size), the probability distribution of the data, embedding, sampling, etc. In the following, we present a critical discussion of the basic requirements for the application of recurrence networks and show that their insufficient application leads to pitfalls in the system-theoretic interpretation of complex network measures.


\paragraph*{Threshold selection.} The crucial algorithmic parameter of recurrence-based time series analysis is $\varepsilon$. Several invariants of a dynamical system (e.g., the 2nd-order R\'enyi entropy $K_2$) can be estimated by taking its recurrence properties for $\varepsilon\to 0$ \cite{Marwan2007}, which suggests that for a feasible analysis of recurrence networks, a low $\varepsilon$ is preferable as well. This is supported by the analogy to complex networks based on spatially extended systems, where attention is usually restricted to the strongest links between individual vertices (i.e., observations from different spatial coordinates) for retrieving meaningful information about relevant aspects of the systems' dynamics \cite{Zhou2006,Donges2009}. In contrast, a high edge density
\begin{equation}
\rho(\varepsilon)=\frac{2E(\varepsilon)}{N(N-1)}
\end{equation}
\noindent
(with $E(\varepsilon)$ being the total number of edges for a chosen $\varepsilon$) does not yield feasible information about the actually relevant structures, because these are hidden in a large set of mainly less important edges.

\begin{figure*}[ht!]
 \centering
 \includegraphics[scale=0.40]{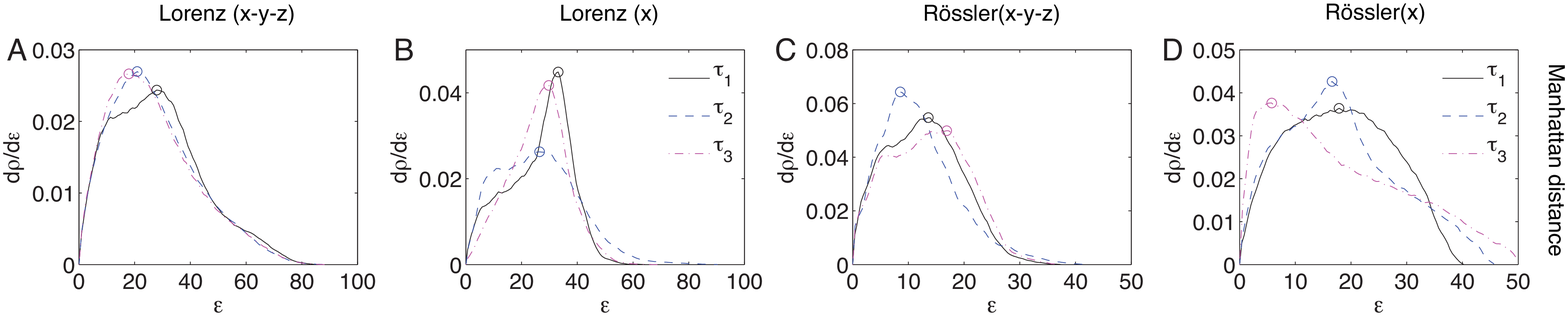}
 \includegraphics[scale=0.40]{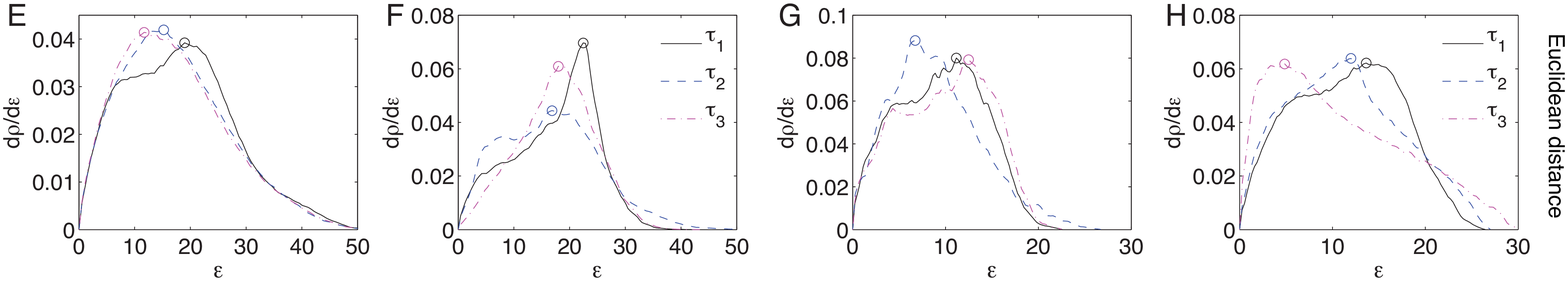}
 \includegraphics[scale=0.40]{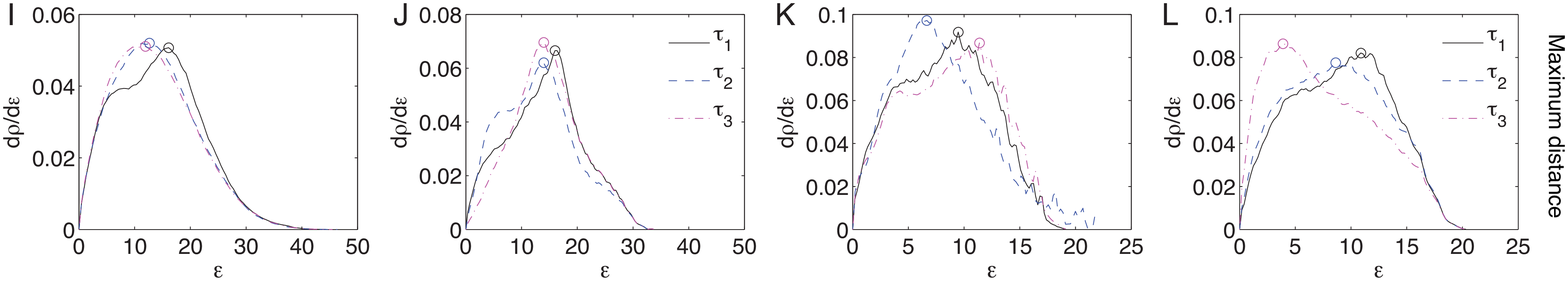}
\caption{\small {(Color online) Effects of different metrics and embeddings on the $\rho(\varepsilon)$ relationship, expressed in terms of the corresponding first derivative. (A,B,C,D): Manhattan distance; (E,F,G,H): Euclidean distance; (I,J,K,L): maximum distance. (A,E,I): Lorenz system $\left( \dot{x}=10(y-x),\ \dot{y}=x(28-z),\ \dot{z}=xy-\frac{8}{3}z\right)$ with original components at three different randomly chosen initial conditions. (B,F,J): Same Lorenz system embedded from the $x$ component with embedding delays $\tau_1=5, \tau_2=15$, and $\tau_3 = 20$. (C,G,K): same as (A,E,I) for the R\"ossler system $\left(\dot{x}=-y-z,\ \dot{y}=x+0.2y,\ \dot{z}=z(x-5.7)\right)$. (D,H,L): same as (B,F,J) for the R\"ossler system with $\tau_1=10, \tau_2=15$ and $\tau_3=20$. Circles indicate the respective maxima. In all cases, time series of $N=1,000$ points with a sampling time of $\Delta t=0.05$ have been used, obtained with a 4th-order Runge-Kutta integrator with fixed step width $h=0.01$. The values of $\tau_2$ are guided by the first zeros of the corresponding auto-correlation functions.} \label{fig1_metric} }
\end{figure*}

As a consequence, only those states should be connected in a recurrence network that are closely neighbored in phase space, leading
to rather sparse networks. Following a corresponding rule of thumb recently confirmed for recurrence quantification analysis
\cite{Schinkel2008}, we suggest choosing $\varepsilon$ as corresponding to an edge density $\rho\lesssim 0.05$
\cite{Marwan2009,Donner2009}, which yields neighborhoods covering appropriately small regions of phase space. Note that since many
topological features of recurrence networks are closely related to the local phase space properties of the underlying
attractor~\cite{Donner2009}, the corresponding information is best preserved for such low $\varepsilon$ unless the presence of
noise requires higher $\varepsilon$ \cite{Schinkel2008}.

Recently, a heuristic criterion has been proposed by Gao and Jin, which selects $\varepsilon$ as the (supposedly unique) turning
point $\varepsilon_{crit}$ in the $\rho(\varepsilon)$ relationship of certain dynamical systems \cite{Gao2009}, formally reading
\begin{equation}
\left. \frac{d\rho}{d\varepsilon}\right|_{\varepsilon=\varepsilon_{crit}}=\max!, \quad \left. \frac{d^2\rho}{d\varepsilon^2}\right|_{\varepsilon=\varepsilon_{crit}}=0.
\label{tpc}
\end{equation}
\noindent
In contrast to our above considerations, for different realizations of the Lorenz system, this turning point criterion yields link
densities of $\rho_{crit}=\rho(\varepsilon=\varepsilon_{crit})\sim 0.15\dots 0.3$ \cite{Gao2009}, implying that considerably large
regions of the attractor are covered by the corresponding neighborhoods. In such cases, it is however \textit{not} possible to
attribute certain network features to specific \textit{small-scale} attractor properties in phase space. More generally,
$\varepsilon$ should be chosen in such a way that small variations in $\varepsilon$ do not induce large variations in the results
of the analysis. In contrast, the turning point criterion (\ref{tpc}) explicitly selects $\varepsilon$ such that small
perturbations in its value will result in a maximum variation of the results. Moreover, besides our general considerations
supporting low $\varepsilon$, application of the turning point criterion leads to serious pitfalls:

(i) $\varepsilon_{crit}$ and, hence, $\rho_{crit}$ depend on the specific metric used for defining distances in phase space
(Fig.~\ref{fig1_metric}). Moreover, experimental time series often contain only a single scalar variable, so that embedding might
be necessary. Since the detailed shape of the attractor in phase space is affected by the embedding parameters, changing the
embedding delay has a substantial effect on $\varepsilon_{crit}$, which is particularly visible in the R\"ossler system (see
Fig.~\ref{fig1_metric} (D,H,L)). { An improper choice of embedding parameters would further increase the variance of $\varepsilon_{crit}$ and will generally not yield meaningful results.} In a similar way, depending on the choice of the other parameters the sampling time of the time
series may also influence the recurrence properties~\cite{Facchini2007} (and, hence, $\varepsilon_{crit}$), since temporal
coarse-graining can cause a loss of detections of recurrences.

\begin{figure}[t!]
 \centering
 \includegraphics[width=\columnwidth]{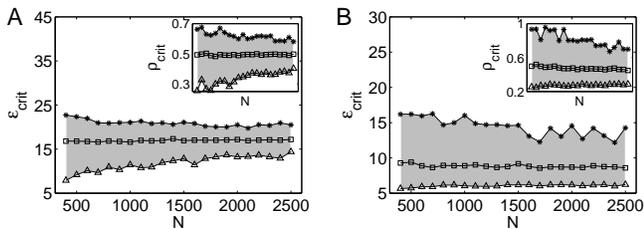}
\caption{Mean values (squares) and range (shaded areas) of turning points $\varepsilon_{crit}$ in 200 independent realizations of
the Lorenz (A) and the R\"ossler system (B) in dependence on the network size $N$ (Euclidean distance, $\Delta t=0.05$). The insets
show the corresponding link densities $\rho_{crit}$ for the same range of $N$.}
\label{lorros_dist_length}
\end{figure}

(ii) The $\varepsilon$-selection should be as independent as possible of the particular realization of the studied system,
especially from the initial conditions and the length $N$ of the time series. The turning point $\varepsilon_{crit}$ after
conditions (\ref{tpc}) is however \textit{not} independent of the specific initial conditions (Fig.~\ref{fig1_metric} (A,E,I) and
(C,G,K)): while its \textit{average} value does not change much with changing $N$, there is a large variance among the individual
trajectories that converges only slowly with increasing $N$ (Fig.~\ref{lorros_dist_length}). Hence, for the same system and the
same network size, already slightly different initial conditions may yield strong differences in $\varepsilon_{crit}$ and
$\rho_{crit}$ (Fig.~\ref{lorros_dist_length}, inset) and, hence, the topological features of the resulting networks.

\begin{figure}[t!]
 \centering
 \includegraphics[width=\columnwidth]{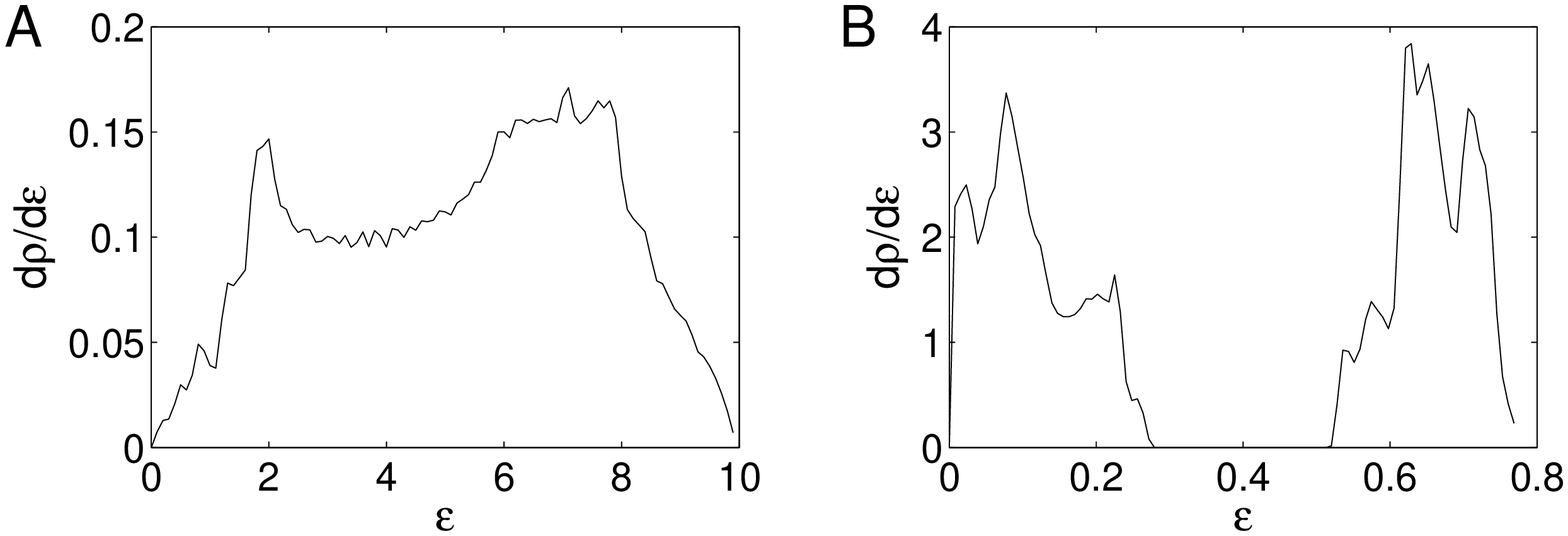}
\caption{Examples for multiple turning points of $d\rho/d\varepsilon$: (A) quasiperiodic trajectory of a continuous system (torus)
and (B) a weakly chaotic orbit of the standard map ($x_{n+1}=x_n+5\sin(y_n) \mod 1$, $y_{n+1}=y_n+x_{n+1} \mod 1$, see
\cite{Zou2007}).}
\label{dist_stdmap}
\end{figure}

(iii) One has to emphasize that the turning point criterion is \textit{not} generally applicable, since there are various typical
examples for both discrete and continuous dynamical systems that are characterized by \textit{several} maxima of
$d\rho(\varepsilon)/d\varepsilon$ (Fig.~\ref{dist_stdmap}).

The above considerations are mainly of concern when studying properties of (known) dynamical systems. In applications to real-world
time series with typically a small number of data or even non-stationarities, it is still possible to derive meaningful
\textit{qualitative} results from small time series networks. However, for a detailed system-theoretic interpretation the use of
smaller recurrence thresholds is recommended~\cite{Marwan2009}.

\paragraph*{Topology of recurrence networks.} The topological features of recurrence networks are closely related to invariant
properties of the observed dynamical system \cite{Marwan2009,Donner2009,Gao2009}. However, a system-theoretic interpretation of the
resulting network characteristics is feasible only based on a careful choice of $\varepsilon$, avoiding the pitfalls outlined
above. For example, many paradigmatic network models as well as real-world systems have been reported to possess small-world
properties (i.e., a high clustering coefficient $\mathcal{C}$ and low average path length $\mathcal{L}$). However, it can be shown
that $\mathcal{C}$ and $\mathcal{L}$ are both functions of $\varepsilon$. In particular, $\mathcal{L}\sim 1/\varepsilon$ (for given
$N$), since spatial distances are approximately conserved in recurrence networks, whereas the specific $\varepsilon$-dependence of
$\mathcal{C}$ varies between different systems.

In addition to the aforementioned global network characteristics, specific vertex properties characterize the local attractor
geometry in phase space in some more detail, where the spatial resolution is determined by $\varepsilon$. In particular, the
\textit{local} clustering coefficient $\mathcal{C}_v$, which quantifies the relative amount of triangles centered at a given vertex
$v$, gives important information about the geometric structure of the attractor within the $\varepsilon$-neighborhood of $v$ in
phase space. Specifically, if the neighboring states form a lower-dimensional subset than the attractor, it is more likely that
closed triangles emerge than for a neighborhood being more uniformly filled with states \cite{Dall2002}. Hence, high values of
$\mathcal{C}_v$ indicate lower-dimensional structures that may correspond to laminar regimes \cite{Marwan2009} or dynamically
invariant objects like unstable periodic orbits (UPOs) \cite{Donner2009}. The relationship with UPOs follows from the fact that
trajectories tend to stay in the vicinity of such orbits for a finite time~\cite{Lathrop_pra_1989}, which leads to a certain amount
of states being accumulated along the UPO with a distinct spatial geometry that differs from that in other parts of a chaotic
attractor. However, since there are infinitely many UPOs embedded in chaotic attractors, such objects (even of a low order) can
hardly be detected using large $\varepsilon$ (where the resulting neighborhoods cover different UPOs) and short time series as
recently suggested \cite{Gao2009}. In contrast, they may be well identified using low $\varepsilon$ and long time series
\cite{Donner2009}.

\begin{figure}[t!]
 \centering
 \includegraphics[width=\columnwidth]{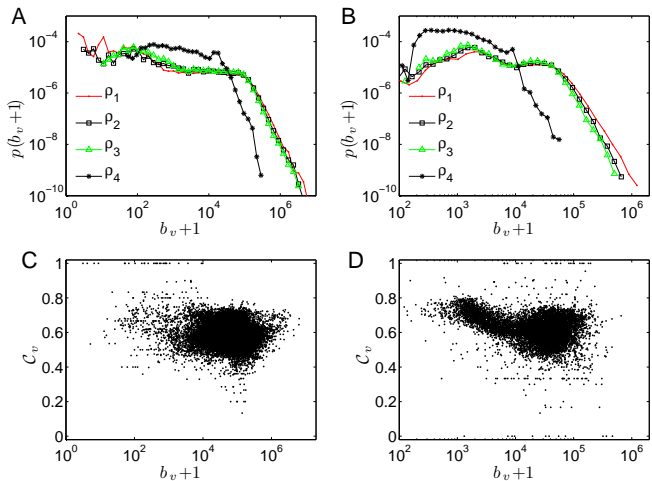}
\caption{(Color online) Probability distribution function of betweenness centrality $b_v$ (in logarithmic scale) for different edge
densities ($\rho_1=0.005$, $\rho_2=0.01$, $\rho_3=0.015$, $\rho_4=0.2$) for the Lorenz (A, $N=20,000$) and R\"ossler system (B,
$N=10,000$), and corresponding relationships between local clustering coefficient $\mathcal{C}_v$ and betweenness centrality $b_v$
(C: Lorenz, D: R\"ossler, $\rho=0.01$) obtained from the original data using the Euclidean distance ($\Delta t=0.05$). }
\label{lorros_cc_bc}
\end{figure}

Another intensively studied vertex property is betweenness centrality $b_v$, which quantifies the relative number of shortest paths
in a network that include a given vertex $v$ \cite{Freeman1979}. In a recurrence network, vertices with high $b_v$ correspond to
regions with low phase space density that are located between higher density regions. Hence, $b_v$ yields information about the
local fragmentation of an attractor. In particular, since phase space regions close to the outer boundaries of the corresponding
attractors do not contribute to many shortest paths, the vertices located in these regions are characterized by low $b_v$, which is
(at least for the Lorenz oscillator) even enhanced by a lower state density. For the sharp inner boundary of the R\"ossler
oscillator, one may observe the opposite behavior. For phase space regions close to low-period UPOs, one also finds lower values of
$b_v$ due to the accumulation of states along these structures (many alternative paths). As the distribution of $b_v$
(Fig.~\ref{lorros_cc_bc} (A,B)) suggests, these features are robust for low $\varepsilon$, but may significantly change if
$\varepsilon$ gets too large (i.e., $\rho=0.2$).

We conclude that in a recurrence network, both $\mathcal{C}_v$ and $b_v$ are sensitive to the presence of UPOs, but resolve
complementary aspects (see Fig.~\ref{lorros_cc_bc}). For the R\"ossler system, we find two distinct maxima in the betweenness
distribution, which are related to the inner and outer parts of the attractor, respectively. In particular, the abundance of low
values is promoted by a high state density at the outer boundary of the attractor near the $x$-$y$ plane, which coincides with a
period-3 UPO \cite{Thiel2003}. In contrast, for the Lorenz system there is no second maximum of $p(b_v)$, since the outer parts
of the attractor are more diffuse and characterized by a considerably lower phase space density than in the R\"ossler attractor. In
both cases, vertices with a high clustering coefficient $\mathcal{C}_v$ are characterized by a broad continuum of betweenness
values, which suggests that $b_v$ is no universal indicator for the presence of UPOs, whereas $\mathcal{C}_v$ allows an approximate
detection of at least low-periodic UPOs in phase space.

In summary, transforming time series into complex networks yields
complementary measures for characterizing phase space properties
of dynamical systems. This work has provided empirical arguments
that the recently suggested approach based on the recurrence
properties in phase space allows a detailed characterization of
dynamically relevant aspects of phase space properties of the
attractor, given that (i) the considered time series is long
enough to be representative for the system's dynamics and (ii) the
threshold distance $\varepsilon$ in phase space for defining a
recurrence is chosen small enough to resolve the scales of
interest. In particular, using the network-theoretic measures
discussed here, the turning point criterion for threshold
selection~\cite{Gao2009} often does \textit{not} allow feasible
conclusions about dynamically relevant structures in phase space.
In contrast, for sufficiently low recurrence thresholds (we
suggest $\rho\lesssim 0.05$ as a rule of thumb), small-scale
structure may be resolved appropriately by complex network
measures, which allow identification of invariant objects such as
UPOs by purely geometric means. { We emphasize that
although our presented considerations have been restricted to
paradigmatic example systems, recurrence networks and related
methods have already been successfully applied to real-world data,
e.g., a paleoclimate record~\cite{Marwan2009} or seismic
activity~\cite{Davidsen2008}. Since these examples are typically
characterized by non-stationarities and non-deterministic
components, we conclude that recurrence networks are promising for
future applied research on various interdisciplinary problems
(e.g.,~\cite{Perc2005}).

As a final remark, we note that the problem of parameter selection
arises for most other network-based methods of time series
analysis (see~\cite{Donner2009} for a detailed comparison).
Important examples include cycle networks~\cite{Zhang2006} with a
correlation threshold, and $k$-nearest neighbor
networks~\cite{Xu2008} with a fixed number $k$ of neighbors as
free parameters, respectively. A general framework for parameter
selection in the context considered here would consequently be
desirable. Other methods are parameter-free, but may suffer from
conceptual limitations and strong intrinsic assumptions. For
example, the currently available visibility graph
concepts~\cite{Lacasa2008} are restricted to univariate time
series.}

\textit{Acknowledgments.} This work has been financially supported
by the German Research Foundation (SFB 555, project C4 and DFG
project no. He 2789/8-2), the Max Planck Society, {the
Leibniz Society (project ECONS),} and the Potsdam Research Cluster
PROGRESS (BMBF). All complex networks measures have been
calculated using the \texttt{igraph} package \cite{Csardi2006}.

\end{document}